\documentclass[10pt,journal]{IEEEtran}

\usepackage{siunitx} 
\usepackage[final]{graphicx} 
\usepackage{amsmath} 
\usepackage{amsthm} 
\usepackage{amssymb,times}
\usepackage{subfigure}
\usepackage{cite}
\usepackage{algorithm}
\usepackage{algorithmic}
\usepackage[english]{babel}
\usepackage{color}
\usepackage{url}
\usepackage{cases}
\usepackage{hyperref}

\newcommand{\ba}{\begin{array}}
\newcommand{\ea}{\end{array}}
\newcommand{\be}{\begin{displaymath}}
\newcommand{\ee}{\end{displaymath}}
\newcommand{\ben}{\begin{equation}}
\newcommand{\een}{\end{equation}}
\newcommand{\bena}{\begin{eqnarray}}
\newcommand{\eena}{\end{eqnarray}}
\newcommand{\beqa}{\begin{eqnarray*}}
\newcommand{\enqa}{\end{eqnarray*}}

\newcommand{\bc}{\begin{center}}
\newcommand{\ec}{\end{center}}
\newcommand{\bi}{\begin{itemize}}
\newcommand{\ei}{\end{itemize}}
\newcommand{\benu}{\begin{enumerate}}
\newcommand{\eenu}{\end{enumerate}}
\newcommand{\bdes}{\begin{description}}
\newcommand{\edes}{\end{description}}
\newcommand{\bt}{\begin{tabular}}
\newcommand{\et}{\end{tabular}}

\newcommand \thetabf{\boldsymbol{\theta}}

\newcommand \xibf{\boldsymbol{\xi}}

\newcommand \Thetabf{\boldsymbol{\Theta}}

\newcommand \abf{{\bf a}}
\newcommand \bbf{{\bf b}}

\newcommand \dbf{{\bf d}}

\newcommand \gbf{{\bf g}}
\newcommand \hbf{{\bf h}}

\newcommand \tbf{{\bf t}}

\newcommand \Abf{{\bf A}}
\newcommand \Bbf{{\bf B}}

\newcommand \Dbf{{\bf D}}

\newcommand \Gbf{{\bf G}}
\newcommand \Hbf{{\bf H}}

\newcommand \Tbf{{\bf T}}

\newcommand{\Cset}{{\mathbb C}}

\newcommand{\circlambda}{\mbox{$\Lambda$
             \kern-.85em\raise1.5ex
             \hbox{$\scriptstyle{\circ}$}}\,}

\begin{document}
\title{Joint Active and Passive Beamforming for IRS-Assisted Radar
\thanks{This work was supported in part by the National Science Foundation under grant
  ECCS-1923739.}
}

\author{Fangzhou Wang,~\IEEEmembership{Student Member,~IEEE,} and
 Hongbin Li,~\IEEEmembership{Fellow,~IEEE,} and Jun Fang,~\IEEEmembership{Senior Member,~IEEE}
\thanks{F. Wang and H. Li are with the Department of Electrical and Computer Engineering, Stevens Institute of Technology, Hoboken, NJ 07030 USA (e-mail: fwang11@stevens.edu; Hongbin.Li@stevens.edu)}
\thanks{J. Fang is with the National Key Laboratory of Science and Technology on Communications, University of Electronic Science and Technology of China, Chengdu 611731, China (e-mail: JunFang@uestc.edu.cn).}
}

\maketitle

\begin{abstract}
Intelligent reflecting surface (IRS) is a promising technology being considered for future wireless communications due to its ability to control signal propagation. This paper considers the joint active and passive beamforming problem for an IRS-assisted radar, where multiple IRSs are deployed to assist the surveillance of multiple targets in cluttered environments. Specifically, we aim to maximize the minimum target illumination power at multiple target locations by jointly optimizing the active beamformer at the radar transmitter and the passive phase-shift matrices at the IRSs, subject to an upperbound on the clutter power at each clutter scatterer. The resulting optimization problem is nonconvex and solved with a sequential optimization procedure along with semedefinite relaxation (SDR). Simulation results show that IRSs can help create effective line-of-sight (LOS) paths and thus substantially improve the radar robustness against target blockage.
\end{abstract}


\IEEEpeerreviewmaketitle

\section{Introduction}
In recent years, intelligent reflecting surface (IRS) has been proposed as a promising technology for reconfiguring the wireless propagation environment via software-controlled reflections \cite{Basar2019,Swindlehurst2021,WangFangLiSPL2020,WangFangLi2020}. IRS is a planar surface comprising a large number of low-cost passive reflecting elements, which are capable of changing the phase and polarization of the impinging signals, thereby collaboratively achieving controllable signal reflection. In contrast to existing wireless link adaptation techniques at the transmitter/receiver, IRS proactively modifies the wireless channel between them and provides an additional degree of freedom to realize a programmable wireless environment \cite{WangFangLiTWC2021}.

Although the use of IRS was firstly proposed for communication purposes, it has recently gained significant attention within the radar research community. One group of works examined the integration of IRS for cooperative radar and communication systems \cite{WangFei21,WangFeiWCL21,JiangRihan21,sankar2021joint}. Specifically, the IRS was employed to mitigate multi-user interference by  joint waveform design and passive beamforming for a dual-functional radar-communication (DFRC) system \cite{WangFei21}, and respectively, spectrum sharing between multi-input multi-output (MIMO) radar and multi-user multi-input single-output communication systems \cite{WangFeiWCL21}. In \cite{JiangRihan21}, the IRS was utilized in a DFRC system to improve target detection in environments with severe path loss, while \cite{sankar2021joint} proposed to adaptively partition the IRS to enhance the radar sensing and communication capabilities of a millimeter-wave DFRC system. Another group of recent works, e.g., \cite{LuDeng21,LuLin21,buzzi2021radar,buzzi2021foundations}, focused on exploiting IRS to enhance the sensing performance in radar. Specifically, the phase-shift matrix of the IRS was optimized for colocated MIMO radar \cite{LuDeng21}, and respectively, distributed MIMO radar \cite{LuLin21} to improve the estimation and detection performance. Target detection was also considered in cases when the radar is aided by a single IRS \cite{buzzi2021radar} or multiple IRSs \cite{buzzi2021foundations}.

Detection and localization of moving targets such as pedestrians and vehicles in urban environments is a challenging radar problem \cite{ThaiBosse2019}. Unlike conventional radar applications where targets are in line of sight (LOS) of the radar, the presence of buildings in urban environments may block the LOS path. One way to handle the problem is to use a radar network. For example, \cite{GuoAdriaan2019} proposed a distributed radar to create continuous coverage in urbanized environments. A different and more economic way is to employ the IRS technology. Along this direction, \cite{aubry2021reconfigurable} analyzed the effect of IRS on radar equations for surveillance in non-line of sight (N-LOS) scenarios.

In this paper, we consider a scenario where multiple IRSs are deployed to assist radar surveillance in a cluttered environment, where the LOS paths from the radar transmitter to prospective targets may be blocked by clutter scatterers. A joint active and passive beamforming design problem is studied, where the minimum target illumination power of multiple targets is maximized by jointly optimizing the transmit beamforming vector and the IRSs' phase-shift matrices under a total transmit power constraint as well as an upperbound on the tolerable clutter power at each clutter scatterer. The proposed design results in a nonconvex constrained optimization problem. We propose a sequential optimization procedure which employs semidefinite relaxation (SDR) to iteratively optimize the active and passive beamformers. Numerical results show that our joint design for the IRS-assisted radar can significantly decrease the probability of blockage for targets over conventional radar system that is not equipped with an IRS, or the IRS-assisted radar system that employs only the active or passive design.
\section{Problem formulation}
\label{sec:model}
Consider an IRS-assisted radar system, where a multi-antenna transmitter tracks multiple targets with the help of multiple IRSs. The transmitter (TX) is equipped with $M$ antenna elements. A beamforming vector $\tbf\in\Cset^{M\times1}$ is designed to transmit a common waveform. Suppose $K$ IRSs are employed to assist the illumination of multiple prospective targets. Each IRS is equipped with $N$ reflecting elements. Each element on the IRS combines all received signals and then re-scatters the combined signal with a certain phase shift \cite{Bjornson2020}. Let $\theta_{k,n}\in[0,~2\pi]$ denote the phase shift associated with the $n$-th element of the $k$-th IRS. The diagonal matrix accounting for the phase response of the $k$-th IRS can be expressed as
\ben
\Thetabf_k\triangleq \text{diag}(e^{\jmath\theta_{k,1}},\cdots,e^{\jmath \theta_{k,N}}).
\een
Suppose there are $L$ targets and $Q$ clutter scatterers located in the surveillance area. Then, the signals from the radar TX and $K$ IRSs superimposed at the $\ell$-th target is given by
\begin{align}
\big(\hbf_{\text{t},\ell}^T+\sum_{k=1}^K\hbf_{\text{i},k,\ell}^T\Thetabf_k^H\Dbf_k\big)\tbf,
\end{align}
and, likewise, the signals from the TX and IRSs superimposed at the $q$-th clutter scatterer is given by
\begin{align}
\big(\gbf_{\text{t},q}^T+\sum_{k=1}^K\gbf_{\text{i},k,q}^T\Thetabf_k^H\Dbf_k\big)\tbf,
\end{align}
where
\begin{itemize}
\item $\hbf_{\text{t},\ell}\in\Cset^{M\times 1}$ denotes the channel from the TX to the $\ell$-th target, $\hbf_{\text{i},k,\ell}\in\Cset^{N\times1}$ the channel from the $k$-th IRS to the $\ell$-th target, and $\Dbf_k\in\Cset^{N\times M}$ the channel from the TX to the $k$-th IRS.
\item $\gbf_{\text{t},q}\in\Cset^{M\times 1}$ and $\gbf_{\text{i},k,q}\in\Cset^{N\times1}$ denote the channel from the TX, and respectively, the $k$-th IRS to the $q$-th clutter scatterer.
\end{itemize}

The problem of interest is to jointly design the transmit beamformer $\tbf$ and the passive diagonal phase-shift matrices $\{\Thetabf_k\}$ by maximizing the minimum illumination power at the target locations, subject to a total radar transmit power constraint as well as an upperbound on the clutter power for each clutter scatterer. Specifically, our joint design can be formulated as the following optimization problem:
\begin{subequations}\label{eq:P1}
\begin{gather}
\max_{\tbf,\{\Thetabf_k\}}~~\min_{\ell=1,\cdots,L}~~\Big\vert\big(\hbf_{\text{t},\ell}^T+\sum_{k=1}^K\hbf_{\text{i},k,\ell}^T\Thetabf_k^H\Dbf_k\big)\tbf\Big\vert^2\\
\text{s.t.}~\tbf^H\tbf\leq \kappa,\\
\Big\vert\big(\gbf_{\text{t},q}^T+\sum_{k=1}^K\gbf_{\text{i},k,q}^T\Thetabf_k^H\Dbf_k\big)\tbf\Big\vert^2\leq\eta_q,~\forall q,\\
\Thetabf_k=\text{diag}(e^{\jmath\theta_{k,1}},\cdots,e^{\jmath \theta_{k,N}}),~\forall k,
\end{gather}
\end{subequations}
where $\kappa$ is the total transmit power and $\eta_q$ is the upperbound of the tolerable clutter power for the $q$-th clutter scatterer.
\section{Proposed Solution}
In this section, a solution to the joint design is presented. Note that problem \eqref{eq:P1} is nonconvex with respect to (w.r.t.) the design variables. However, if we solve the problem iteratively by fixing either $\tbf$ or $\Thetabf_k$ from the last iteration, the original problem can be decomposed into two simpler subproblems. In other words, a sequential optimization procedure can be employed to solve the max-min problem \eqref{eq:P1} by iteratively optimizing the worst-case illuminated power at the target locations w.r.t. the transmit beamformer $\tbf$ and passive beamformers $\Thetabf_k$. Specifically, by fixing $\Thetabf_k$ to the values obtained from the $j$-th iteration $\Thetabf_k^{(j)}$, we can write \eqref{eq:P1} as
\begin{subequations}\label{eq:P_transmit}
\begin{gather}
\max_{\tbf}~~\min_{\ell=1,\cdots,L}~~\Big\vert\big(\hbf_{\text{t},\ell}^T+\sum_{k=1}^K\hbf_{\text{i},k,\ell}^T(\Thetabf_k^{(j)})^H\Dbf_k\big)\tbf\Big\vert^2\\
\text{s.t.}~\tbf^H\tbf\leq\kappa,\label{eq:sdrcon1}\\
\Big\vert\big(\gbf_{\text{t},q}^T+\sum_{k=1}^K\gbf_{\text{i},k,q}^T(\Thetabf_k^{(j)})^H\Dbf_k\big)\tbf\Big\vert^2\leq\eta_q,~\forall q\label{eq:sdrcon2}.
\end{gather}
\end{subequations}
This is a nonconvex quadratically constrained quadratic programming (QCQP) problem and can be solved with the semidefinite relaxation (SDR) technique. Specifically, by letting $\Tbf=\tbf\tbf^H$ and dropping the rank-one constraint, problem \eqref{eq:P_transmit} can be rewritten as
\begin{subequations}\label{eq:P_transmit_1}
\begin{gather}
\max_{\Tbf}~~\min_{\ell=1,\cdots,L}~~\text{tr}(\Abf_\ell\Tbf)\\
\text{s.t.}~\text{tr}(\Tbf)\leq\kappa,\\
\text{tr}(\Bbf_q\Tbf)\leq\eta_q,~\forall q,
\end{gather}
\end{subequations}
where $\Abf_\ell=\abf_\ell\abf_\ell^{H}$, $\abf_\ell^H=\hbf_{\text{t},\ell}^T+\sum_{k=1}^K\hbf_{\text{i},k,\ell}^T(\Thetabf_k^{(j)})^H\Dbf_k$, $\Bbf_q=\bbf_q\bbf_q^{H}$, and $\bbf_q^H=\gbf_{\text{t},q}^T+\sum_{k=1}^K\gbf_{\text{i},k,q}^T(\Thetabf_k^{(j)})^H\Dbf_k$.
The above problem is convex and can be solved by standard numerical solvers, e.g., CVX \cite{cvxBoyd2014}.

When employing the SDR approach, one has to convert the solution $\Tbf^{(j+1)}$ to \eqref{eq:P_transmit_1} into a feasible solution $\tbf^{(j+1)}$ to \eqref{eq:P_transmit}. We can use the randomization method to obtain a solution $\tbf^{(j+1)}$ from $\Tbf^{(j+1)}$ \cite{Luozhiquan2010,WangLi2020}. Specifically, given $\Tbf^{(j+1)}$, we can generate a set of independent and identically distributed Gaussian random vectors $\xibf_i\sim\mathcal{CN}(\mathbf{0},\Tbf^{(j+1)})$, $i=1,\cdots,I$, where $I$ is the number of randomization trials. Note that the random vectors $\xibf_i$ are not always feasible for \eqref{eq:P_transmit}, but we can apply a scaling to turn them into feasible solutions. Specifically, $\xibf_i$ can be normalized w.r.t. the largest value of $\big\{\xibf_i^H\xibf_i/\kappa,~\xibf_i^H\Bbf_1\xibf_i/\eta_1,~\cdots,~\xibf_i^H\Bbf_Q\xibf_i/\eta_Q\big\}$, i.e.,
\ben\label{eq:randomization}
\widetilde{\xibf}_i=\frac{\xibf_i}{\sqrt{\max \big\{\xibf_i^H\xibf_i/\kappa,~\xibf_i^H\Bbf_1\xibf_i/\eta_1,~\cdots,~\xibf_i^H\Bbf_Q\xibf_i/\eta_Q\big\}}}.
\een
After the randomization procedure, a feasible rank-one solution is obtained as $\tbf^{(j+1)}=\arg~\max_{\widetilde{\xi}_i}~\min_{\ell=1,\cdots,L}~~\widetilde{\xi}_i^H\Abf_\ell\widetilde{\xi}_i$.

Next, we find $\Thetabf_k$ by fixing $\tbf$ to the value obtained from the latest updates, $\tbf^{(j+1)}$, in which case the optimization problem \eqref{eq:P1} becomes
\begin{subequations}\label{eq:P_passive}
\begin{gather}
\max_{\{\Thetabf_k\}}~~\min_{\ell=1,\cdots,L}~~\Big\vert\big(\hbf_{\text{t},\ell}^T+\sum_{k=1}^K\hbf_{\text{i},k,\ell}^T\Thetabf_k^H\Dbf_k\big)\tbf^{(j+1)}\Big\vert^2\\
\text{s.t.}~
\Big\vert\big(\gbf_{\text{t},q}^T+\sum_{k=1}^K\gbf_{\text{i},k,q}^T\Thetabf_k^H\Dbf_k\big)\tbf^{(j+1)}\Big\vert^2\leq\eta_q,~\forall q,\\
\Thetabf_k=\text{diag}(e^{\jmath\theta_{k,1}},\cdots,e^{\jmath \theta_{k,N}}),~\forall k.
\end{gather}
\end{subequations}
Let $\widetilde{\hbf}_{\text{i},k,\ell}\triangleq\hbf_{\text{i},k,\ell}\circ(\Dbf_k\tbf^{(j+1)})$, $\widetilde{\gbf}_{\text{i},k,q}\triangleq\gbf_{\text{i},k,q}\circ(\Dbf_k\tbf^{(j+1)})$, $\thetabf_k=[e^{\jmath\theta_{k,1}},\cdots,e^{\jmath \theta_{k,N}}]^T$, where $\circ$ denotes the Hadamard (elementwise) product. The above optimization problem can be rewritten as
\begin{subequations}\label{eq:P_passive_1}
\begin{gather}
\max_{\{\thetabf_k\}}~~\min_{\ell=1,\cdots,L}~~\Big\vert\hbf_{\text{t},\ell}^T\tbf^{(j+1)}+\sum_{k=1}^K\thetabf_k^H\widetilde{\hbf}_{\text{i},k,\ell}\Big\vert^2\\
\text{s.t.}~
\Big\vert\gbf_{\text{t},q}^T\tbf^{(j+1)}+\sum_{k=1}^K\thetabf_k^H\widetilde{\gbf}_{\text{i},k,q}\Big\vert^2\leq\eta_q,~\forall q,\\
\vert\thetabf_k(n)\vert=1,~\forall k,~\forall n.
\end{gather}
\end{subequations}
To write the functions in a more compact form, we define $\hbf_{\ell}=[\widetilde{\hbf}_{\text{i},1,\ell}^T,\cdots,\widetilde{\hbf}_{\text{i},K,\ell}^T]^T$, $\gbf_{q}=[\widetilde{\gbf}_{\text{i},1,q}^T,\cdots,\widetilde{\gbf}_{\text{i},K,q}^T]^T$, and $\thetabf=[\thetabf_1^T,\cdots,\thetabf_K^T]^T$. Hence, \eqref{eq:P_passive_1} can be simplified as
\begin{subequations}\label{eq:P_passive_2}
\begin{gather}
\max_{\thetabf}~~\min_{\ell=1,\cdots,L}~~\Big\vert a_\ell+\thetabf^H\hbf_{\ell}\Big\vert^2\\
\text{s.t.}~
\Big\vert b_q+\thetabf^H\gbf_{q}\Big\vert^2\leq\eta_q,~\forall q,\label{eq:constr}\\
\vert\thetabf(l)\vert=1,~l=1,\cdots,NK,
\end{gather}
\end{subequations}
where $a_\ell=\hbf_{\text{t},\ell}^T\tbf^{(j+1)}$ and $b_q=\gbf_{\text{t},q}^T\tbf^{(j+1)}$. Due to the unit modulus constraint placed on the entries of $\thetabf$, the above optimization problem \eqref{eq:P_passive_2} is nonconvex. However, by observing that the objective function and constraint \eqref{eq:constr} can be transformed into quadratic forms, we can apply the SDR technique to approximately solve \eqref{eq:P_passive_2} efficiently.

Specifically, the non-homogeneous QCQP problem \eqref{eq:P_passive_2} can be reformulated as a homogeneous one by introducing an auxiliary variable $t$:
\begin{subequations}\label{eq:P_passive_3}
\begin{gather}
\max_{\bar{\thetabf}}~~\min_{\ell=1,\cdots,L}~~\bar{\thetabf}^H\Hbf_\ell\bar{\thetabf}+\vert a_\ell\vert^2\\
\text{s.t.}~
\bar{\thetabf}^H\Gbf_{q}\bar{\thetabf}+\vert b_q\vert^2\leq\eta_q,~\forall q,\label{eq:cons_clutter}\\
\vert\bar{\thetabf}(l)\vert=1,~l=1,\cdots,NK+1\label{eq:cons_circl},
\end{gather}
\end{subequations}
where $\bar{\thetabf}=[\thetabf^T,t]^T$ and
\begin{equation}
\begin{split}
\Hbf_\ell=
\begin{bmatrix}
\hbf_\ell\hbf_\ell^H& \hbf_\ell a_\ell^\ast\\
\hbf_\ell^H a_\ell& 0
\end{bmatrix},~
\Gbf_q=
\begin{bmatrix}
\gbf_q\gbf_q^H& \gbf_q b_q^\ast\\
\gbf_q^H b_q& 0
\end{bmatrix}.
\end{split}
\end{equation}
Similarly, problem \eqref{eq:P_passive_3} can be solved through the SDR technique. Specifically, by letting $\Thetabf=\bar{\thetabf}\bar{\thetabf}^H$ and dropping the rank-one constraint, problem \eqref{eq:P_passive_3} can be rewritten as
\begin{subequations}\label{eq:P_passive_4}
\begin{gather}
\max_{\Thetabf}~~\min_{\ell=1,\cdots,L}~~\text{tr}(\Thetabf\Hbf_\ell)+\vert a_\ell\vert^2\\
\text{s.t.}~
\text{tr}(\Thetabf\Gbf_{q})+\vert b_q\vert^2\leq\eta_q,~\forall q,\\
\vert\Thetabf(l,l)\vert=1,~l=1,\cdots,NK+1.
\end{gather}
\end{subequations}
It can be observed that problem \eqref{eq:P_passive_4} is a standard convex semidefinite program and can be solved by CVX. A similar randomization procedure as in \eqref{eq:randomization} can be employed to obtain a solution $\thetabf^{(j+1)}$ from $\Thetabf^{(j+1)}$ except that the scaling becomes
\ben\label{eq:randomizationtheta}
\widetilde{\xibf}_i=\frac{\xibf_i}{\sqrt{\max_{q=1,\cdots,Q}~\xibf_i^H\Gbf_q\xibf_i/(\eta_q-\vert b_q\vert^2)}},
\een
which makes $\widetilde{\xibf}_i$ to satisfy the constraint \eqref{eq:cons_clutter}. In addition, $\widetilde{\xibf}_i$ should meet the unit modulus constraint \eqref{eq:cons_circl}, which means the final solution can be recovered by
\ben
\hat{\xibf}_i=e^{\jmath\arg(\frac{\widetilde{\xibf}_i}{\widetilde{\xibf}_i(NK+1)})},
\een
where $\widetilde{\xibf}_i(NK+1)$ denotes the $(NK+1)$-st element of $\widetilde{\xibf}_i$. Then, a feasible rank-one solution is obtained as $\bar{\thetabf}^{(j+1)}=\arg~\max_{\hat{\xi}_i}~\min_{\ell=1,\cdots,L}~~\hat{\xi}_i^H\Hbf_\ell\hat{\xi}_i+\vert a_{\ell}\vert^2$. The alternating process is repeated until the algorithm converges, e.g., the minimum received signal power improvement is smaller than a tolerance $\epsilon$. Our proposed sequential optimization algorithm for the joint design problem is summarized in $\textbf{Algorithm~\ref{alg:sequential}}$.

Note that the computational complexity of the proposed sequential optimization algorithm mainly depends on the number of iterations $J$ and the number of randomization trials for the semidefinite relaxation $I$. On one hand, the two convex problems are solved in each iteration with a complexity of $\mathcal{O}(2JN^{3.5})$ if an interior-point method is used \cite{Boyd2004}. On the other hand, the computational complexities of $I$ randomization trials are in the order of $\mathcal{O}(IM^2)$ for \eqref{eq:randomization}, and respectively, $\mathcal{O}(I(NK+1)^2)$ for \eqref{eq:randomizationtheta}. Thus, the overall complexity of the proposed alternating algorithm is $\mathcal{O}(2JN^{3.5})+\mathcal{O}(IM^2)+\mathcal{O}(I(NK+1)^2)$. Numerical results show that $\textbf{Algorithm~\ref{alg:sequential}}$ usually converges in less than 20 iterations and a sufficiently large number of randomization, e.g., $I=5000$, is required to yield a good solution.

\begin{algorithm}
\caption{Sequential Optimization Algorithm for the Joint Design Formulation in \eqref{eq:P1}}
\begin{algorithmic}
\label{alg:sequential}
\STATE \textbf{Input:} Channel information $\hbf_{\text{t},\ell}$, $\hbf_{\text{i},k,\ell}$, $\hbf_{\text{t},\ell}$, $\hbf_{\text{i},k,\ell}$, $\Dbf_k$, $\kappa$, $\eta_q$, and tolerance $\epsilon$.
\STATE \textbf{Output:} Transmit beamformer $\tbf$ and phase shift matrices $\Thetabf_k$.\\
\STATE  \textbf{Initialization:} Initialize $\Thetabf_k^{(0)}$, and set iteration index $j=0$.\\
\REPEAT
\STATE
\begin{enumerate}

  \item Fix $\Thetabf_k^{(j)}$. Use \eqref{eq:P_transmit_1} and randomization \eqref{eq:randomization} to obtain $\tbf^{(j+1)}$.
  \item Fix $\tbf^{(j+1)}$. Use \eqref{eq:P_passive_4} along with randomization \eqref{eq:randomizationtheta} to find $\thetabf^{(j+1)}$.
  \item Use $\thetabf^{(j+1)}$ to formulate $\Thetabf_k^{(j+1)}$.
         \item Set $j=j+1$.
\end{enumerate}
\UNTIL convergence.
\RETURN $\tbf=\tbf^{(j+1)}$ and $\Thetabf_k=\Thetabf_k^{(j+1)}$.
\end{algorithmic}
\end{algorithm}

\section{Numerical Results}

\begin{figure}[t!]
\centering
\includegraphics[width=2.39in]{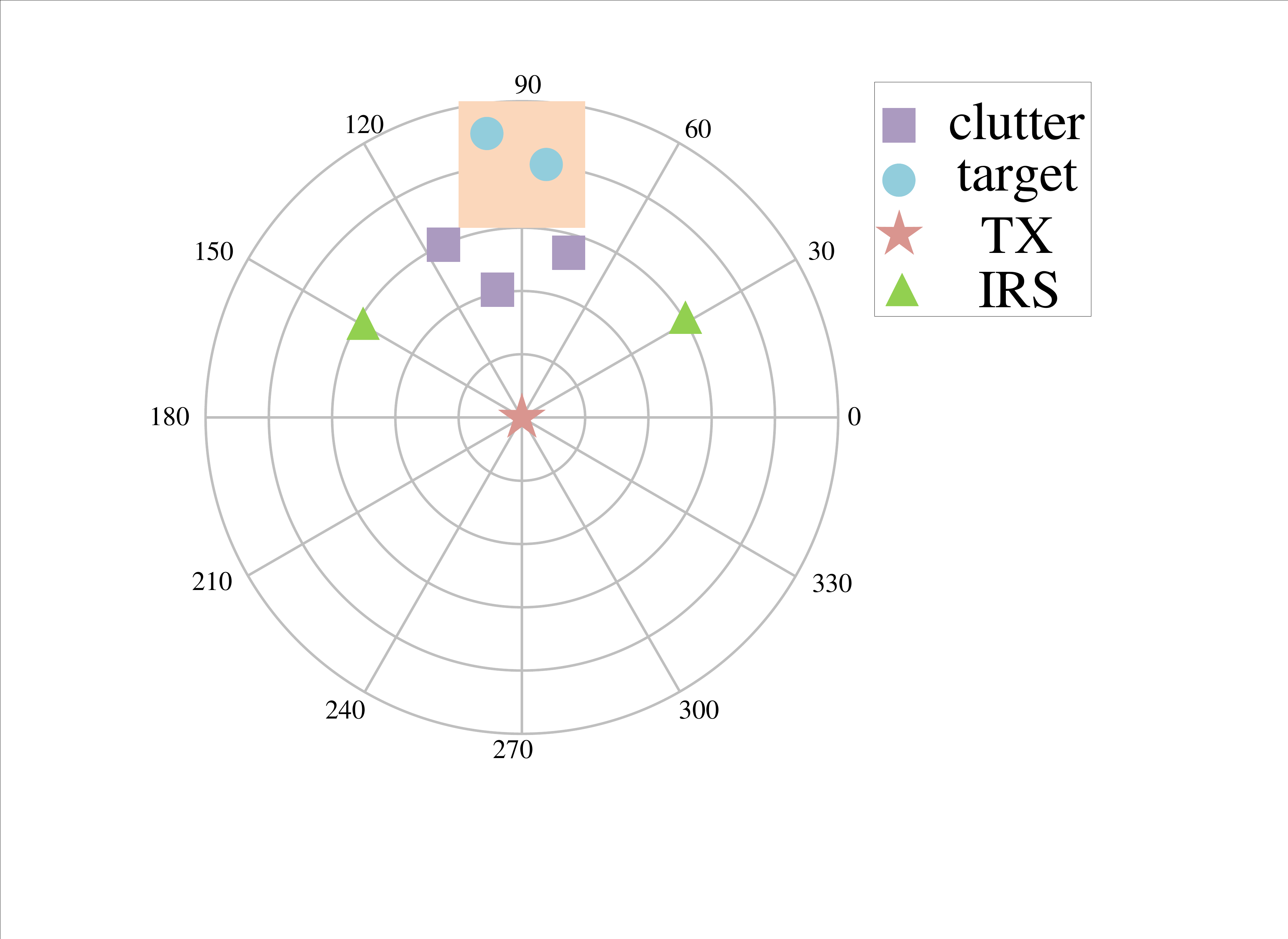}
\caption{Simulation setup.}
\label{fgr:simulationconfiguration}
\end{figure}

In this section, we present numerical results to illustrate the performance of the proposed IRS-assisted scheme. We consider a scenario where the transmitter employs a uniform linear array (ULA) with $M=64$ antennas and each IRS consists of a ULA with $N=100$ reflecting elements. The channel vectors $\hbf_{\text{t},\ell}$, $\hbf_{\text{i},k,\ell}$, $\hbf_{\text{t},\ell}$, and $\hbf_{\text{i},k,\ell}$ are generated according to the following geometric channel model \cite{AyachRajagopal2014} (here we take $\hbf_{\text{t},\ell}$ as an example and the other channel vectors are similarly generated): $\hbf_{\text{t},\ell}=\sqrt{M}\alpha_{\text{t},\ell}\gamma \hbf(\phi_{\text{t},\ell})$, where $\alpha_{\text{t},\ell}$ is the complex gain associated with the path from the TX to the $\ell$-th target, $\phi_{\text{t},\ell}$ is the associated angle of departure, $\hbf\in\Cset^{M\times1}$ represents the normalized array response vector, and $\gamma\in\{1,0\}$ is a binary variable to specify if this path is blocked or not (more discussion on this in the next paragraph). The complex gain is generated according to a complex Gaussian distribution \cite{Akdeniz2014}: $\alpha_{\text{t},\ell}\sim\mathcal{CN}(0,10^{-0.1\rho})$ with $\rho$ given as $\rho=a+10b\log_{10}(d)+\xi$ where $d$ denotes the distance between the transmitter and the target, and $\xi\sim\mathcal{N}(0,\sigma^2_{\xi})$. The values of $a$, $b$, and $\sigma_\xi$ are set to be $a=64$, $b=2$, and $\sigma_\xi=5.8$dB, as suggested by real-world LOS channel measurements \cite{Akdeniz2014}. In addition, the channel matrix between the transmitter and the $k$-th IRS is characterized by $\Dbf_k=\sqrt{M_\text{t}N}\alpha_k\dbf_{\text{r},k}\dbf_{\text{t},k}^H$, where $\alpha_k$ is the complex gain and can be similarly generated as $\alpha_{\text{t},\ell}$, $\dbf_{\text{r},k}$ and $\dbf_{\text{t},k}$ are the normalized array response vectors.

The simulation configuration is shown in Fig.\,\ref{fgr:simulationconfiguration}, where the $(x,y)$-coordinates of the TX are $(0,0)$ and those of the two IRSs are $(-130\,\text{m},75\,\text{m})$ and $(130\,\text{m},75\,\text{m})$. The locations of the targets are randomly distributed within a square box specified by $x\in[-75\,\text{m} ~75\,\text{m}]$ and $y\in[150\,\text{m} ~250\,\text{m}]$. To simulate the effect of blockage, we use the 3dB beamwidth of the TX \cite{Richards05}. Specifically, if the TX beam simultaneously covers multiple objects (targets or clutter scatterers) within the 3dB beamwidth, then the near object will block the far object, which is simulated by setting $\gamma=0$ for the far path. Other system parameters are set as follows: $\epsilon=10^{-3}$, $\eta_q=0.5\mu$W, $\forall q$, $I=5000$. All results are averaged over 100 random channel realizations (also the randomization of the target locations).

In the simulation, the performance of the following four design approaches are included: The \emph{joint design}, \emph{passive design}, and \emph{active design} are the proposed joint active and passive beamforming (design $\tbf$ and $\Thetabf_k$), the passive-only beamforming (design $\Thetabf_k$ by fixing $\tbf$ so that the TX illuminates toward the first IRS), and respectively, the active-only beamforming (design $\tbf$ by fixing $\Thetabf_k$ as identity matrix) for the IRS-assisted radar system; \emph{without IRS} is the conventional radar system that has the TX but no IRS (design $\tbf$).

\begin{figure}[tbh!]
\centering
\includegraphics[width=2.39in]{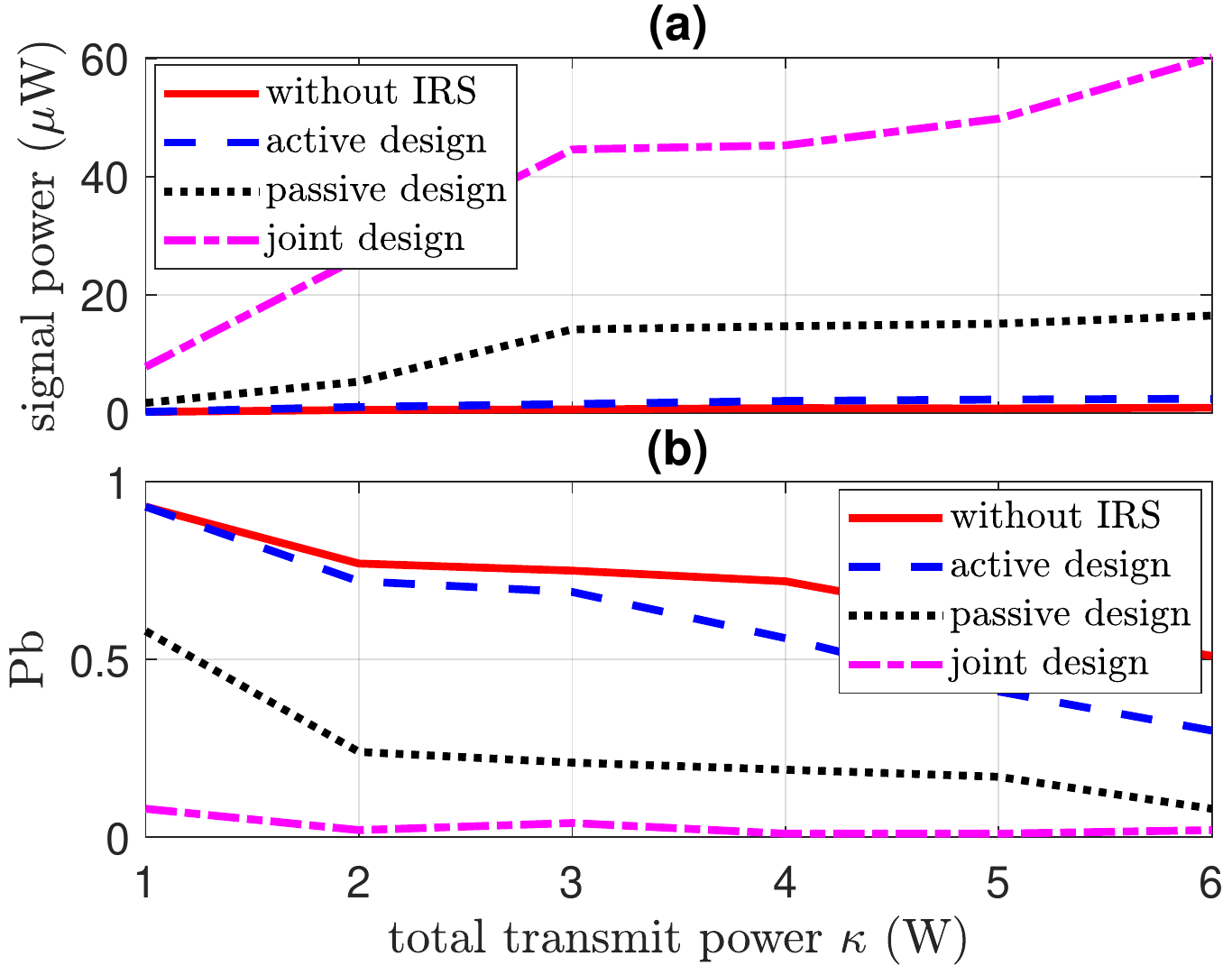}
\caption{Performance versus $\kappa$ when $Q=3$: (a) minimum target illumination power, (b) probability of target blockage.}
\label{fgr:Q3}
\end{figure}

Fig.\,\ref{fgr:Q3} depicts the performance for the four design approaches versus the total transmit power $\kappa$ when $Q=3$ and the clutter scatterers locate at $(-75\,\text{m},125\,\text{m})$, $(0,125\,\text{m})$, and $(75\,\text{m},125\,\text{m})$, respectively. In Fig.\,\ref{fgr:Q3}(a), which shows the minimum target illumination power, it is seen that the three IRS-assisted radars outperform the conventional radar system because the IRSs can help create effective LOS paths to illuminate the targets and thus substantially improve the radar's robustness against blockage. This can also been seen in Fig.\,\ref{fgr:Q3}(b) where the probability of blockage (Pb) is plotted, where blockage occurs when the target illumination power is sufficiently small. In our simulation, the blockage power level is set to $0.5\mu$W. It is observed that although also assisted by IRSs, the active- and passive-only designs are significantly less effective than the joint design.

\begin{figure}[tbh!]
\centering
\includegraphics[width=2.39in]{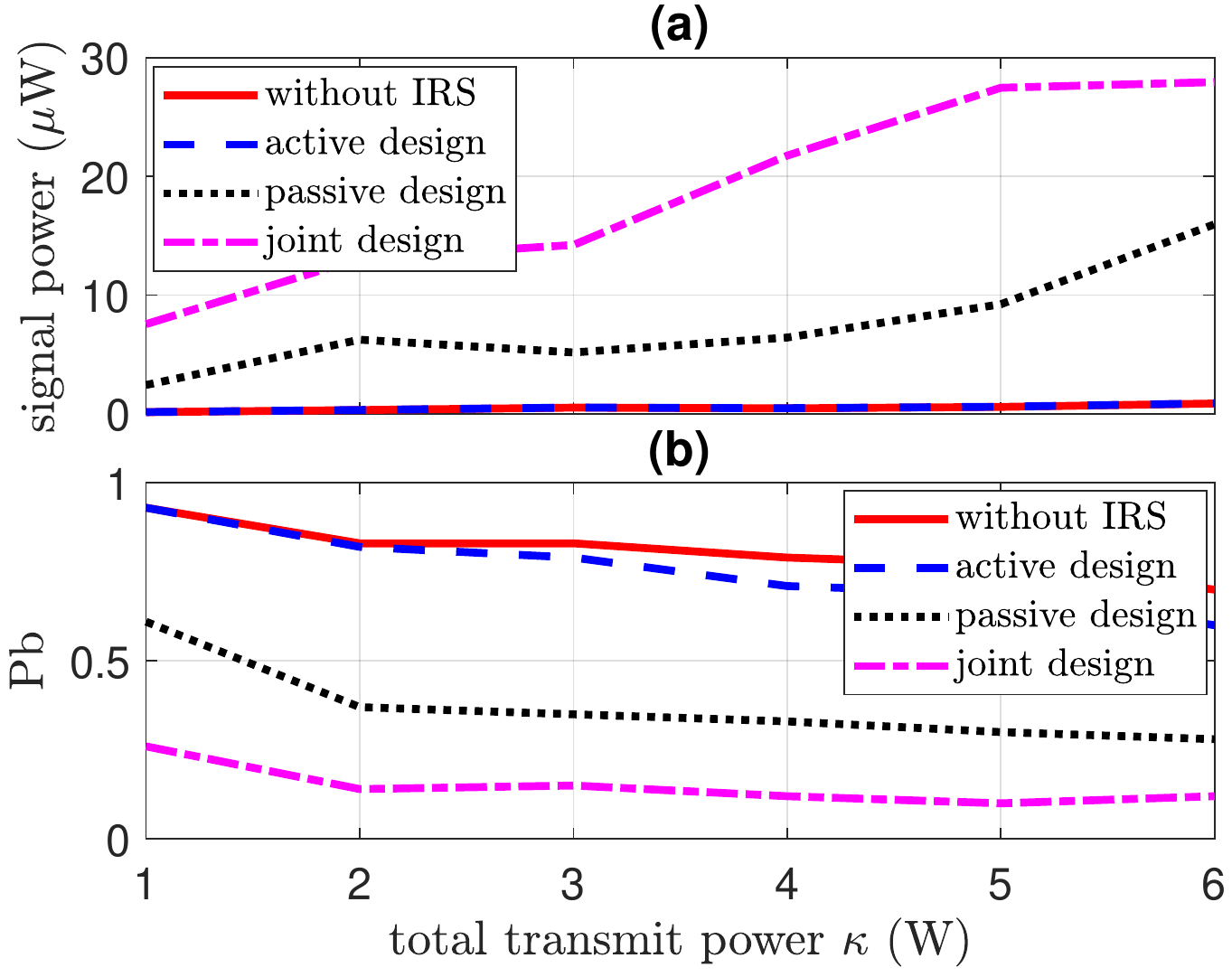}
\caption{Performance versus $\kappa$ when $Q=9$: (a) minimum target illumination power, (b) probability of target blockage.}
\label{fgr:Q9}
\end{figure}

Fig.\,\ref{fgr:Q9} depicts the results for a more densely cluttered environment with $Q=9$ clutter scatterers, where the clutter scatterer locations are $(-75\,\text{m},100\,\text{m})$, $(0,100\,\text{m})$, $(75\,\text{m},100\,\text{m})$, $(-75\,m,125\,m)$, $(0,125\,\text{m})$, $(75\,\text{m},125\,\text{m})$, $(-75\,\text{m},150\,\text{m})$, $(0,150\,\text{m})$, and $(75\,\text{m},150\,\text{m})$. Similar behaviors among the four designs can be observed except that the performance of all decreases. This is because more clutters means that the targets have a higher possibility of being blocked, which reduces the propagation paths. However, the joint design for the IRS-assisted system still outperforms the other designs.

\section{Conclusion}
We proposed an approach to reducing the target blockage in radar systems in cluttered environments by leveraging passive IRSs. Under a total transmit power constraint as well as an upperbound on the tolerable clutter power at each clutter scatterer, the active transmit beamforming vector at the radar and the passive phase-shift matrices at the IRSs were jointly optimized to maximize the minimum target illumination power at multiple target locations. The resulting nonconvex problem was solved by applying the alternating optimization and SDR techniques. It was shown that the IRS-assisted radar system is more robust against target blockage over conventional radar system.
\bibliographystyle{IEEEtran}
\bibliography{IRS}

\begin{thebibliography}{10}
\providecommand{\url}[1]{#1}
\csname url@samestyle\endcsname
\providecommand{\newblock}{\relax}
\providecommand{\bibinfo}[2]{#2}
\providecommand{\BIBentrySTDinterwordspacing}{\spaceskip=0pt\relax}
\providecommand{\BIBentryALTinterwordstretchfactor}{4}
\providecommand{\BIBentryALTinterwordspacing}{\spaceskip=\fontdimen2\font plus
\BIBentryALTinterwordstretchfactor\fontdimen3\font minus
  \fontdimen4\font\relax}
\providecommand{\BIBforeignlanguage}[2]{{%
\expandafter\ifx\csname l@#1\endcsname\relax
\typeout{** WARNING: IEEEtran.bst: No hyphenation pattern has been}%
\typeout{** loaded for the language `#1'. Using the pattern for}%
\typeout{** the default language instead.}%
\else
\language=\csname l@#1\endcsname
\fi
#2}}
\providecommand{\BIBdecl}{\relax}
\BIBdecl

\bibitem{Basar2019}
E.~Basar, M.~Di~Renzo, J.~De~Rosny, M.~Debbah, M.-S. Alouini, and R.~Zhang,
  ``Wireless communications through reconfigurable intelligent surfaces,''
  \emph{IEEE Access}, vol.~7, pp. 116\,753--116\,773, 2019.

\bibitem{Swindlehurst2021}
C.~Pan, H.~Ren, K.~Wang, J.~F. Kolb, M.~Elkashlan, M.~Chen, M.~Di~Renzo,
  Y.~Hao, J.~Wang, A.~L. Swindlehurst, X.~You, and L.~Hanzo, ``Reconfigurable
  intelligent surfaces for {6G} systems: Principles, applications, and research
  directions,'' \emph{IEEE Communications Magazine}, vol.~59, no.~6, pp.
  14--20, 2021.

\bibitem{WangFangLiSPL2020}
P.~Wang, J.~Fang, H.~Duan, and H.~Li, ``Compressed channel estimation for
  intelligent reflecting surface-assisted millimeter wave systems,'' \emph{IEEE
  Signal Processing Letters}, vol.~27, pp. 905--909, 2020.

\bibitem{WangFangLi2020}
P.~Wang, J.~Fang, X.~Yuan, Z.~Chen, and H.~Li, ``Intelligent reflecting
  surface-assisted millimeter wave communications: Joint active and passive
  precoding design,'' \emph{IEEE Transactions on Vehicular Technology},
  vol.~69, no.~12, pp. 14\,960--14\,973, 2020.

\bibitem{WangFangLiTWC2021}
P.~Wang, J.~Fang, L.~Dai, and H.~Li, ``Joint transceiver and large intelligent
  surface design for massive {MIMO} mmwave systems,'' \emph{IEEE Transactions
  on Wireless Communications}, vol.~20, no.~2, pp. 1052--1064, 2021.

\bibitem{WangFei21}
X.~Wang, Z.~Fei, Z.~Zheng, and J.~Guo, ``Joint waveform design and passive
  beamforming for {RIS}-assisted dual-functional radar-communication system,''
  \emph{IEEE Transactions on Vehicular Technology}, pp. 1--1, 2021.

\bibitem{WangFeiWCL21}
X.~Wang, Z.~Fei, J.~Guo, Z.~Zheng, and B.~Li, ``{RIS}-assisted spectrum sharing
  between {MIMO} radar and {MU-MISO} communication systems,'' \emph{IEEE
  Wireless Communications Letters}, vol.~10, no.~3, pp. 594--598, 2021.

\bibitem{JiangRihan21}
Z.-M. Jiang, M.~Rihan, P.~Zhang, L.~Huang, Q.~Deng, J.~Zhang, and E.~M.
  Mohamed, ``Intelligent reflecting surface aided dual-function radar and
  communication system,'' \emph{IEEE Systems Journal}, pp. 1--12, 2021.

\bibitem{sankar2021joint}
\BIBentryALTinterwordspacing
R.~S.~P. Sankar, B.~Deepak, and S.~P. Chepuri, ``Joint communication and radar
  sensing with reconfigurable intelligent surfaces,'' 2021. [Online].
  Available: \url{https://arxiv.org/abs/2105.01966}
\BIBentrySTDinterwordspacing

\bibitem{LuDeng21}
W.~Lu, B.~Deng, Q.~Fang, X.~Wen, and S.~Peng, ``Intelligent reflecting
  surface-enhanced target detection in {MIMO} radar,'' \emph{IEEE Sensors
  Letters}, vol.~5, no.~2, pp. 1--4, 2021.

\bibitem{LuLin21}
W.~Lu, Q.~Lin, N.~Song, Q.~Fang, X.~Hua, and B.~Deng, ``Target detection in
  intelligent reflecting surface aided distributed {MIMO} radar systems,''
  \emph{IEEE Sensors Letters}, vol.~5, no.~3, pp. 1--4, 2021.

\bibitem{buzzi2021radar}
S.~Buzzi, E.~Grossi, M.~Lops, and L.~Venturino, ``Radar target detection aided
  by reconfigurable intelligent surfaces,'' \emph{IEEE Signal Processing
  Letters}, vol.~28, pp. 1315--1319, 2021.

\bibitem{buzzi2021foundations}
\BIBentryALTinterwordspacing
S.~Buzzi, E.~Grossi, M.~Lops, and L.~Venturino, ``Foundations of {MIMO} radar detection aided by reconfigurable
  intelligent surfaces,'' 2021. [Online]. Available:
  \url{https://arxiv.org/abs/2105.09250}
\BIBentrySTDinterwordspacing

\bibitem{ThaiBosse2019}
K.-P.-H. Thai, O.~Rabaste, J.~Bosse, D.~Poullin, I.~D.~H. Sáenz, T.~Letertre,
  and T.~Chonavel, ``Detection-localization algorithms in the around-the-corner
  radar problem,'' \emph{IEEE Transactions on Aerospace and Electronic
  Systems}, vol.~55, no.~6, pp. 2658--2673, 2019.

\bibitem{GuoAdriaan2019}
X.~Guo, C.~S. Ng, E.~de~Jong, and A.~B. Smits, ``Concept of distributed radar
  system for mini-{UAV} detection in dense urban environment,'' in \emph{2019
  International Radar Conference (RADAR)}, 2019, pp. 1--4.

\bibitem{aubry2021reconfigurable}
\BIBentryALTinterwordspacing
A.~Aubry, A.~D. Maio, and M.~Rosamilia, ``Reconfigurable intelligent surfaces
  for {N-LOS} radar surveillance,'' 2021. [Online]. Available:
  \url{https://arxiv.org/abs/2104.00456}
\BIBentrySTDinterwordspacing

\bibitem{Bjornson2020}
E.~Bj$\ddot{\text{o}}$rnson, d.~$\ddot{\text{O}}$zdogan, and E.~G. Larsson,
  ``Reconfigurable intelligent surfaces: Three myths and two critical
  questions,'' \emph{IEEE Communications Magazine}, vol.~58, no.~12, pp.
  90--96, 2020.

\bibitem{cvxBoyd2014}
M.~Grant and S.~Boyd, ``{CVX}: Matlab software for disciplined convex
  programming, version 2.1,'' \url{http://cvxr.com/cvx}, Mar. 2014.

\bibitem{Luozhiquan2010}
Z.-Q. Luo, W.-K. Ma, A.~M.-C. So, Y.~Ye, and S.~Zhang, ``Semidefinite
  relaxation of quadratic optimization problems,'' \emph{IEEE Signal Processing
  Magazine}, vol.~27, no.~3, pp. 20--34, 2010.

\bibitem{WangLi2020}
F.~Wang and H.~Li, ``Joint waveform and receiver design for co-channel hybrid
  active-passive sensing with timing uncertainty,'' \emph{IEEE Transactions on
  Signal Processing}, vol.~68, pp. 466--477, 2020.

\bibitem{Boyd2004}
S.~Boyd and L.~Vandenberghe, \emph{Convex Optimization}.\hskip 1em plus 0.5em
  minus 0.4em\relax Cambridge University Press, 2004.

\bibitem{AyachRajagopal2014}
O.~E. Ayach, S.~Rajagopal, S.~Abu-Surra, Z.~Pi, and R.~W. Heath, ``Spatially
  sparse precoding in millimeter wave {MIMO} systems,'' \emph{IEEE Transactions
  on Wireless Communications}, vol.~13, no.~3, pp. 1499--1513, 2014.

\bibitem{Akdeniz2014}
M.~R. Akdeniz, Y.~Liu, M.~K. Samimi, S.~Sun, S.~Rangan, T.~S. Rappaport, and
  E.~Erkip, ``Millimeter wave channel modeling and cellular capacity
  evaluation,'' \emph{IEEE Journal on Selected Areas in Communications},
  vol.~32, no.~6, pp. 1164--1179, 2014.

\bibitem{Richards05}
M.~A. Richards, \emph{Fundamentals of Radar Signal Processing}.\hskip 1em plus
  0.5em minus 0.4em\relax New York, NY, USA: McGraw-Hill, 2005.

\end{thebibliography}
\end{document}